\documentclass[%
reprint,
 amsmath,amssymb,
prb,
]{revtex4-2}

\usepackage{graphicx}
\usepackage{dcolumn}
\usepackage{bm}
\usepackage{xcolor}
\usepackage{placeins}
\usepackage{setspace}
\usepackage{bbold}

\newcommand{\be}{\begin{equation}}
\newcommand{\ee}{\end{equation}}
\newcommand{\bfig}{\begin{figure}}
\newcommand{\efig}{\end{figure}}

\usepackage{lineno}
\usepackage{lipsum}

\begin{document}      

\title{Eavesdropping on competing condensates by the edge supercurrent in a Weyl superconductor
}

\author{Stephan Kim$^{1,2}$}
\author{Shiming Lei$^3$}
\author{Leslie M. Schoop$^3$}
\author{R. J. Cava$^3$}
\author{N. P. Ong$^{1,\S}$}
\affiliation{
{Departments of Physics$^1$, Electrical and Computer Engineering$^2$ and Chemistry$^3$, Princeton University, Princeton, NJ 08544, USA}
}

\date{\today}      

\maketitle     
\noindent

\noindent
{\bf 
In a topological insulator the metallic surface states are easily distinguished from the insulating bulk states~\cite{FuKane07}. By contrast, in a topological superconductor~\cite{FuKane08,Qi,FuBerg,Oppen}, much less is known about the relationship between an edge supercurrent and the bulk pair condensate. Can we force their pairing symmetries to be incompatible?
In the superconducting state of the Weyl semimetal MoTe$_2$, an edge supercurrent is observed as oscillations in the current-voltage (\emph{I-V}) curves induced by fluxoid quantization~\cite{Wang}. We have found that the $s$-wave pairing potential of supercurrent injected from niobium contacts is incompatible with the intrinsic pair condensate in MoTe$_2$. The incompatibility leads to strong stochasticity in the switching current $I_c$ as well as other anomalous properties such as an unusual antihysteretic behavior of the ``wrong'' sign. Under supercurrent injection, the fluxoid-induced edge oscillations survive to much higher magnetic fields \emph{H}. Interestingly, the oscillations are either very noisy or noise-free depending on the pair potential that ends up dictating the edge pairing. Using the phase noise as a sensitive probe that eavesdrops on the competiting bulk states, we uncover an underlying blockade mechanism whereby the intrinsic condensate can pre-emptively block proximitization by the Nb pair potential depending on the history. 
}

\vspace{3mm}\noindent
\emph{Preliminaries}\\\noindent
As reported in Ref.~\cite{Wang}, the edge supercurrent in MoTe$_2$ realizes a toroidal topology despite the absence of holes in the exfoliated crystals. As $H$ is slowly increased, fluxoid quantization leads to a sawtooth field profile for the edge superfluid velocity ${\bf v}_s$, which translates to oscillations of $I_{\rm c}$ that are clearly seen in a color-map plot of $dV/dI$ vs. $H$ (a fluxoid is a flux quantum $\phi_0$ plus the superfluid circulation; Sec. VI in Supplementary Materials SM). The large number of oscillation cycles ($\sim 90$) allow its phase noise to be analyzed. The analysis reveals that a large phase noise appears whenever edge-state pairing, driven by Nb, is incompatible with bulk-state pairing in MoTe$_2$. Incompatibility between the injected $s$-wave supercurrent and paired states intrinsic to MoTe$_2$ has observable consequences in both the bulk and edge states. By varying the contact geometry across 8 devices, we can distinguish edge-state features from those in the bulk.

Because the interfaces between Nb and MoTe$_2$ have high transmittivity at 20 mK, our experiment lies in the strong-coupling proximity-effect regime, well beyond the Josephson effect regime (Sec. V in SM). In strong coupling junctions, the system (Nb plus MoTe$_2$) is treated as a single superconductor with a unique $T_c$ and a gap function $\hat\Psi({\bf r})$ that is inhomogeneous~\cite{deGennes,deGennesbook,deGennesDeutscher,Rainer,Sauls,Sigrist}. In device $SA$, $T_c$ ($\simeq 850$ mK) lies well above the unperturbed $T_c$ of pristine MoTe$_2$ (100 mK) although still far below that of Nb ($\sim 8$ K). Similarly, the critical field and critical current are greatly enhanced over the values in pristine MoTe$_2$.

To discuss the proximity effect (Sec. VII of SM)~\cite{deGennes,deGennesbook,deGennesDeutscher,Rainer,Sauls,Sigrist}, we denote the Gor'kov pair amplitude in Nb by $\eta({\bf r}) = \langle \hat{\psi}_\downarrow({\bf r})\hat{\psi}_\uparrow({\bf r})\rangle$ where $\hat{\psi}_\alpha({\bf r})$ annihilates an electron in spin state $\alpha=\uparrow,\downarrow$ at $\bf r$. We call the pair amplitude in pristine MoTe$_2$ (of unknown symmetry) $F_{\alpha\beta}({\bf r}) =\langle \hat{\psi}_\alpha({\bf r})\hat{\psi}_\beta({\bf r})\rangle$. 
Aside from MoTe$_2$, evidence for edge supercurrents has been reported in Bi nanowires~\cite{Bouchiat} and in a Kagome superconductor~\cite{Ali}.

\begin{figure*}[t]
\includegraphics[width=18 cm]{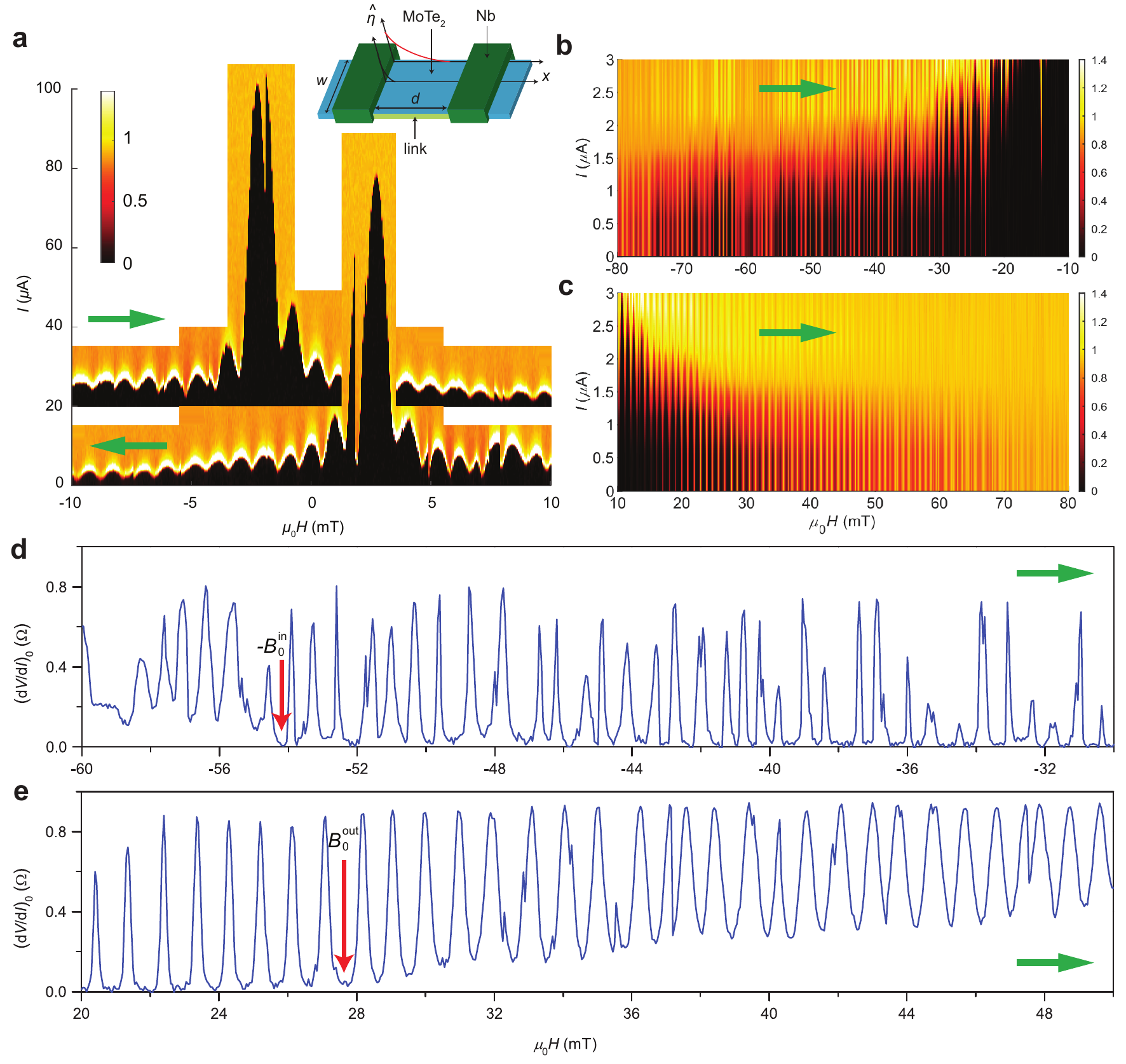}
\caption{\label{figcentral} 
Anti-hysteretic central peak and edge oscillations observed in MoTe$_2$ measured by supercurrent injection from Nb pads in device $SA$ at 18 mK. Panel (a) shows the color map of the differential resistance $dV/dI$ plotted versus magnetic field $H$ and current $I$. Dissipationless regions are shown in black (color scale at left). The central peak occurs at -3 or +3 mT when $H$ is scanned from left to right or right to left, respectively. The inset defines the width $w$ and spacing $d$ of the device. The decays of the Nb pairing amplitudes in the bulk and along an edge are sketched. Panels (b) and (c) display the fluxoid-induced edge oscillations on the inbound ($d|H|/dt<0$) and outbound branches ($d|H|/dt>0$), respectively.  
Panels (d) and (e) plot the zero-bias differential resistance $(dV/dI)_0$ vs. $H$ on the inbound and outbound branches, respectively. The phase noise is much larger in (d) than in (e). The transition to dissipationless behavior ($(dV/dI)_0\to 0$) occurs at $-B^{\rm in}_0 = -54$ mT and $B^{\rm out}_0 = +28$ mT in (d) and (e), respectively (red arrows). In each panel, the green arrow indicates field scan direction.
}
\end{figure*}

\vspace{3mm}\noindent
\emph{Anti-hysteretic behavior of central peak}\\\noindent
Figure \ref{figcentral}a shows the color map of the differential resistance $dV/dI$ versus $H$ and current $I$ measured in device $SA$ at temperature $T = 18$ mK. In weak $H$, supercurrent injection from Nb leads to a large critical current $I_{\rm c}$ within a narrow central peak (shown shaded black) that reaches values of $\sim 80\; \mu$A (20$\times$ larger than seen in pristine crystals). An immediate surprise is the anomalous position of the central peak. In conventional superconductors, we expect $H$ to cross zero before the flux density $B$ does. Instead, the central peak here emerges before $H$ reaches zero for either scan (which seems oddly anti-causal). We dub this curious pattern ``anti-hysteretic.''

When $|H|$ exceeds 10 mT, we observe a dense array of periodic peaks in the color map (Figs. \ref{figcentral}b and \ref{figcentral}c). These are the fluxoid-induced oscillations~\cite{Wang}, now persisting to 90 mT (compared with 3 mT in pristine crystals). As shown in Figs. \ref{figcentral}d and \ref{figcentral}e, the oscillations appear as narrow peaks in the differential resistance $(dV/dI)_0$ measured at $I=0$. At each peak, ${\bf v}_s$ reverses direction to admit a fluxoid~\cite{Wang}. 

\begin{figure*}[t]
\includegraphics[width=15 cm]{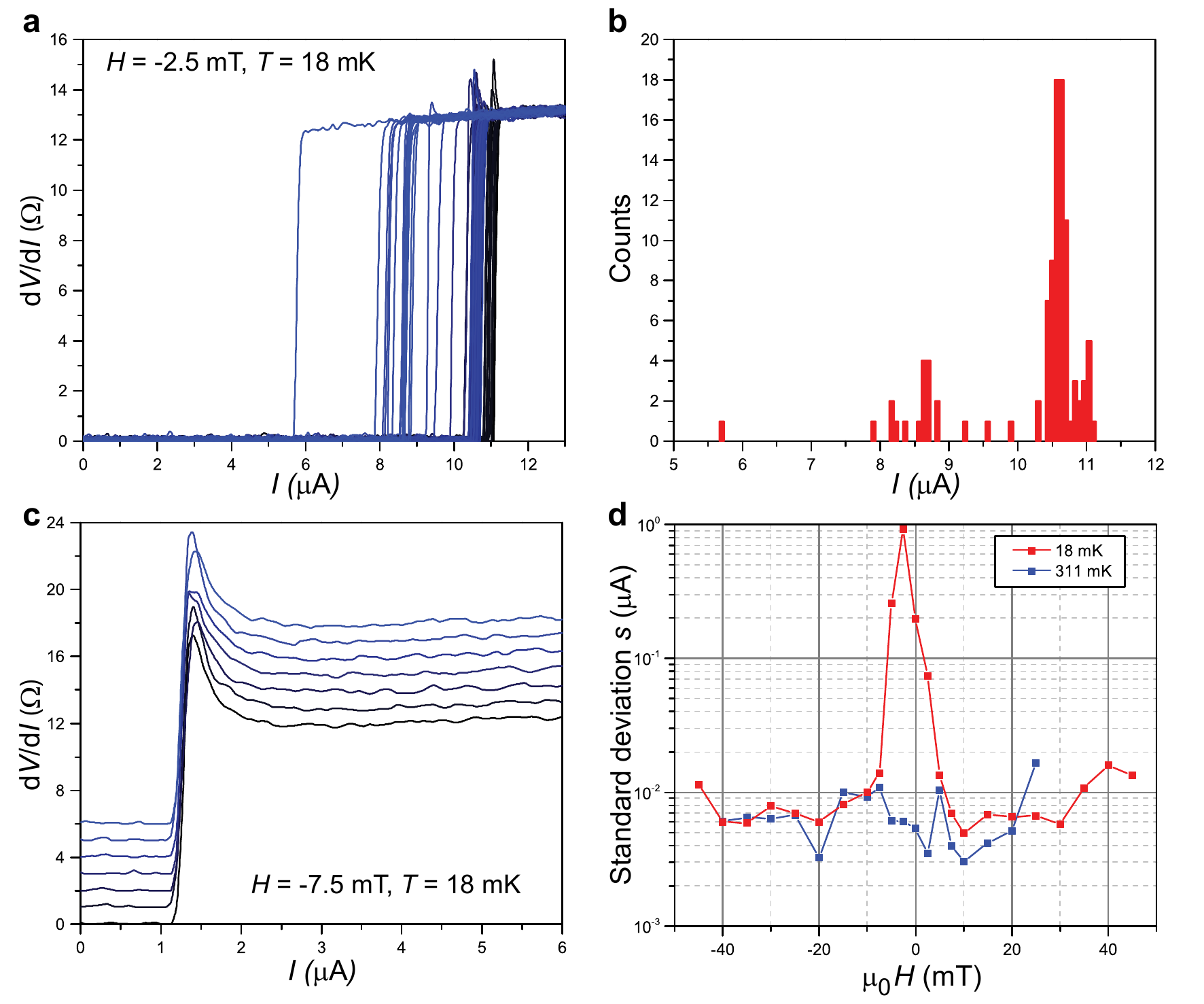}
\caption{\label{figstochastic} 
Stochastic switching in bulk states measured in device SC. When $H$ lies within the central peak ($H$ = $-2.5$ mT, Panel a), the switch from dissipationless to dissipative behavior is stochastic. The critical current in 100 curves of $dV/dI$ vs. $I$ shows bunching of $I_c$ suggestive of a bimodal distribution $P(I_{\rm c}, H)$. In Panel b, the histogram plot of $P(I_{\rm c}, H)$ confirms the bimodal distribution.
However, when $H$ lies outside the central peak ($-7.5$ mT), the switching transitions are non-stochastic. In all 100 scans the transition occurs at $1.27\pm 0.01\;\mu$A (7 are shown in Panel c). Panel d plots the standard deviation $s$ vs. $H$ of the distribution $P(I_{\rm c}, H)$ at 18 mK (Sec. II in SM). At the peak, $s$ is a 100$\times$ larger than its value outside the peak. At 311 mK, the peak is unresolved. 
}
\end{figure*}

\vspace{3mm}\noindent
\emph{Stochastic Switching}\\\noindent
The incompatibility between the intrinsic pair condensation amplitude $F_{\alpha\beta}$ and $\eta$ injected from Nb is most evident in $I$-$V$ curves measured when $H$ lies within the central peak in Fig. \ref{figcentral}a. The switching transition is strongly stochastic (Sec. II in SM). In Fig. \ref{figstochastic}a, we plot 100 curves of $dV/dI$ vs. $I$ measured at 18 mK in device $SC$ with $H$ fixed at $-2.5$ mT. The critical currents $\{I_{\rm c}\}$ obey a distribution function $P(I_{\rm c},H)$ that is distinctly bimodal, with a histogram that shows a major peak at $10.5 \;\mu$A and a minor one at 8.5 $\mu$A (see Fig. \ref{figstochastic}b and Sec. II in SM). In the color map of $dV/dI(H,I)$ (Fig.  \ref{figcentral}a), the stochastic nature of the distribution is also apparent as random spikes within the central peak. 
By contrast, outside the central peak ($H = -7.5$ mT), the distribution $P(I_{\rm c},H)$ becomes very narrow (Fig. \ref{figstochastic}c). The standard deviation $s(H)$ (obtained from 100 scans at each $H$) is plotted in Fig. \ref{figstochastic}d. The profile $s(H)$ displays a narrow peak at the central peak with peak value 100$\times$ larger than the baseline (red curve). At 311 mK, however, the peak is absent (blue curve).
The stochastic switching reflects the competition between bulk-date pairing with symmetry $\eta$ or $F_{\alpha\beta}$.

\vspace{3mm}\noindent
\emph{Phase Noise and Anti-hysteretic $(dV/dI)_0$ Curves}\\\noindent
The competition between $F_{\alpha\beta}$ and $\eta$ has a strong effect on the phase noise of the oscillations in $(dV/dI)_0$. Field scans are called inbound if $|H|$ decreases with time, and outbound if $|H|$ increases. 
On inbound branches (Figs. \ref{figcentral}b and \ref{figcentral}d), the noise content of the oscillations is invariably very large, whereas, on outbound branches (Figs. \ref{figcentral}c and \ref{figcentral}e), the phase noise is conspicuously suppressed, especially in the interval $5<H<45$ mT, which we call the ``quiet'' zone.

\begin{figure*}[t]
\includegraphics[width=18 cm]{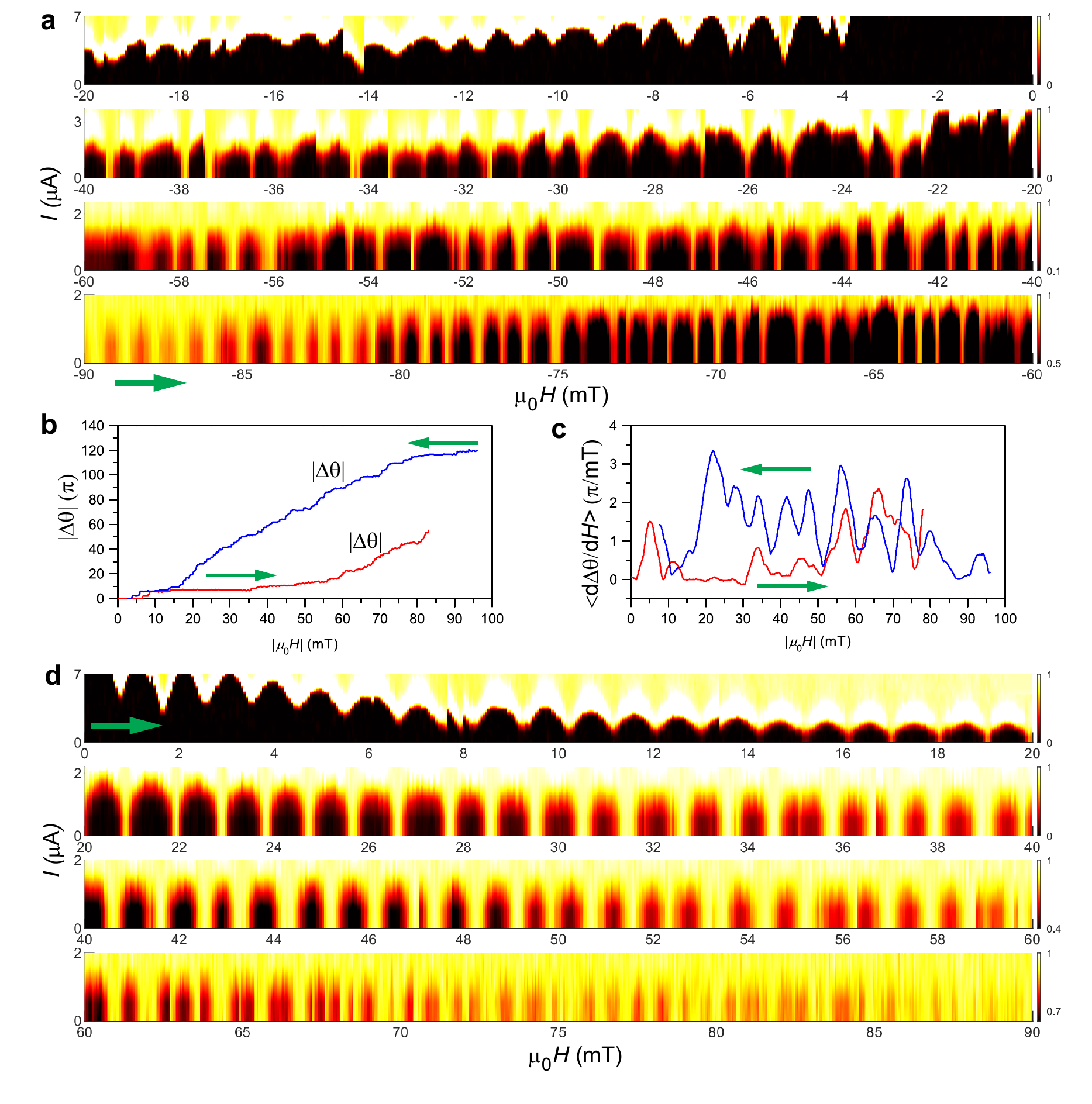}
\caption{\label{fignoise}
Nature of the phase noise in the fluxoid-induced oscillations over the full field interval in Sample $SA$ at 18 mK. The four strips in Panel (a) show the observed color map of $dV/dI$ in the $H$-$I$ plane as $H$ is scanned from -90 mT to 0 mT (inbound branch). Initially, ($-90<H<-80$ mT) the phase noise is quite small. Between -80 mT and -12 mT, the emergence of large phase noise strongly distorts the oscillations. The noise is dominated by 2$\pi$ jumps in the phase $\theta(H)$ which lead to random compressions and dilations of the period. We attribute the noise to incompatibility between edge pairing induced by $\eta$ (Nb) and the intrinisic pair amplitude $F_{\alpha\beta}$ in the vortex-liquid state of MoTe$_2$. Panel (b) compares the curves of the phase deviation $|\Delta\theta(H)|$ in the inbound branch (blue) with the outbound (red). $\Delta\theta(H) = \theta(H)-\theta_0(H)$ measures the phase deviation caused by random phase jumps accumulated over the entire field interval (see text). Panel (c) compares the smoothed derivative $\langle d\Delta\theta/dH\rangle$ between the inbound (blue) and outbound branches (red). The large phase noise in the inbound branch causes $\langle d\Delta\theta/dH\rangle$ to lie well above the outbound, especially between 5 and 45 mT. The four strips in Panel (d) show the color map of $dV/dI$ as $H$ is swept from 0 to 90 mT (outbound branch). Over the ``quiet'' zone (5 to 45 mT), the phase noise is negligible except for isolated 2$\pi$ jumps (at 13.4, 35 and 36.5 mT). In all panels the green arrow indicates field scan direction.
}
\end{figure*}

In Fig. \ref{fignoise} we emphasize the differences by displaying the color map of the oscillations over the entire field range. On the inbound branch (Fig. \ref{fignoise}a), the noise content is much higher than on the outbound (Fig. \ref{fignoise}d). To quantify the phase noise, we express the fluctuations in the frequency as an $H$-dependent phase $\theta(H)$ (Sec. III in SM). It is expedient to subtract from $\theta(H)$ a background $\theta_0(H)$ derived from an $H$-linear fit to $\theta(H)$ in the quiet zone (5, 45) mT (Sec. III in SM). The deviation $\Delta\theta(H)=\theta(H)-\theta_0(H)$ then highlights field intervals with large phase noise.
In Fig. \ref{fignoise}b, $\Delta\theta$ on the inbound branch (blue curve) is seen to be much larger than on the outbound curve (red). In the quiet zone (5, 45) mT, $\Delta\theta$ is flat by design. We also employ the smoothed derivative $\langle d\Delta\theta/dH\rangle$ which is more sensitive to local changes in the frequency. As seen in Fig. \ref{fignoise}c, $\langle d\Delta\theta/dH\rangle$ is much larger on the inbound branch than in the quiet zone on the outbound.

\begin{figure*}[t]
\includegraphics[width=18 cm]{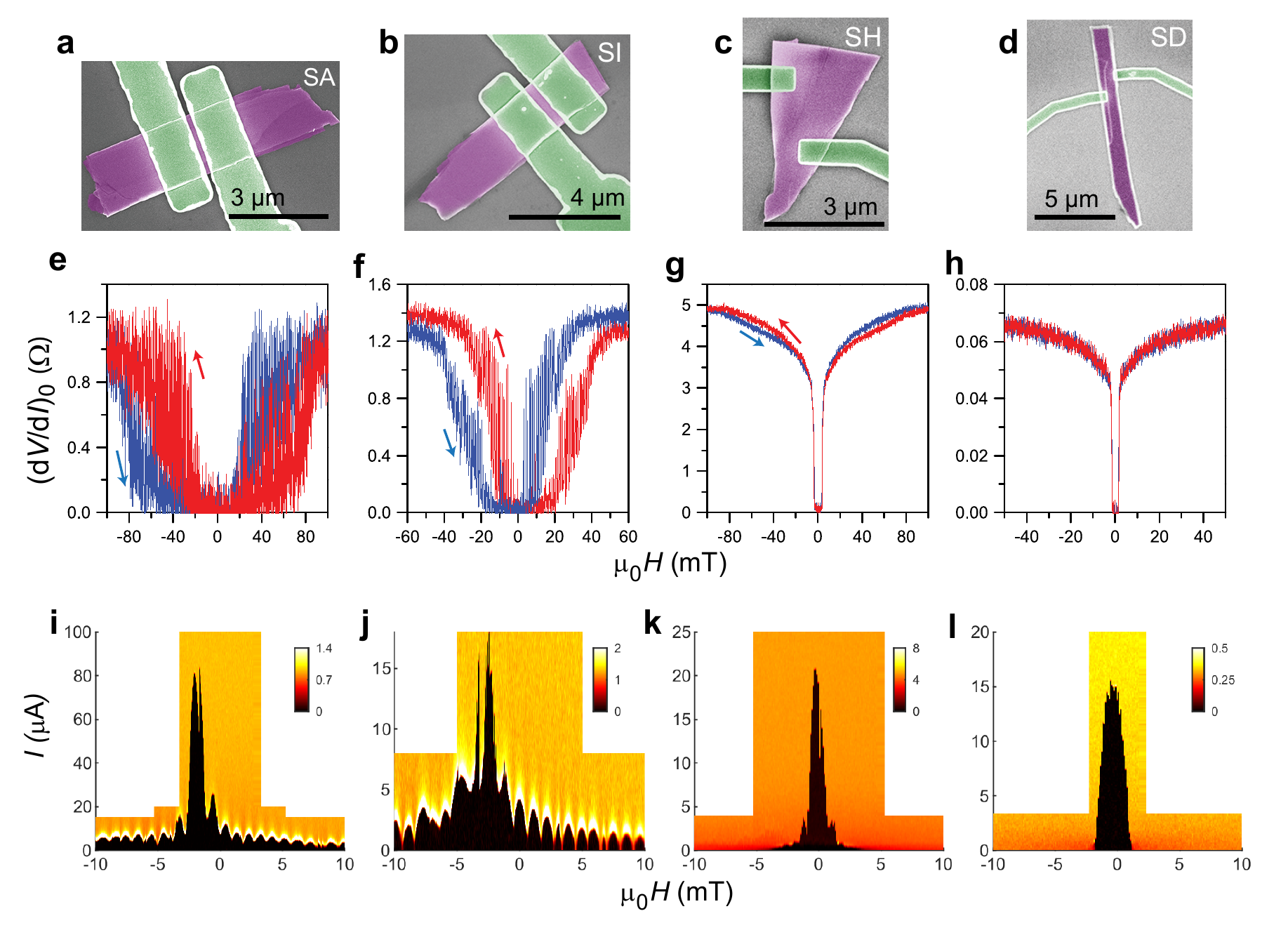}
\caption{\label{figsamples} 
Systematic suppression of edge oscillations and anti-hysteresis in four devices ($SA, SI, SH$ and $SD$ in Panels a,..., d, respectively). $SA$ and $SI$ (with spacing $d$ = 156 and 285 nm, respectively) are examples from a series of 5 parallel-strip devices. $SH$ and $SD$ are from a second series in which the Nb contact volumes are decreased while $\ell_1$ and $\ell_2$ are progressively increased. 
The middle row (Panels e,...,h) shows the corresponding antihystereses in $(dV/dI)_0$ vs. $H$ (blue and red arrows indicate field scan directions). The corresponding color maps of $dV/dI$ vs. $H$ and $I$ are in the bottom row (i,..., l). In devices $SA$ and $SI$, prominent antihysteresis and edge oscillations are seen, both in $(dV/dI)_0$ (Panels e, f) and their color maps (i, j). In devices $SH$ and $SD$, these group II features are suppressed or absent in $(dV/dI)_0$ (Panels g and h) and in the color maps (k and l). See Table S1 for parameters in the 8 devices and Fig. S3 for results from devices $SK,SJ,SB,SF$.
}
\end{figure*}

Aside from the oscillations, $(dV/dI)_0$ also displays a dissipationless regime bracketed by the two fields $B^{\rm in}_0$ and $B^{\rm out}_0$ (red arrows in Figs. \ref{figcentral}d and \ref{figcentral}e). An expanded view of these transitions is shown in Figs. \ref{figsamples}a -- \ref{figsamples}h for 4 devices. In device $SA$, the transition on the inbound branch occurs at $-B^{\rm in}_0 = -54$ mT, larger in magnitude than the field $B^{\rm out}_0$ (= 28 mT) on the outbound branch (blue curves in Fig. \ref{figsamples}e). The reverse sweep (red curves) is the mirror reflection of the blue curve. The resulting hysteresis is again anti-hysteretic.

\vspace{3mm}\noindent
\emph{Distinguishing Edge from Bulk Supercurrents}\\\noindent
The observed supercurrents segregate into two groups. In group I (central peak in Fig. \ref{figcentral}a), $I_{\rm c}$ attains very large values ($80-200\;\mu$A) but is easily suppressed in a weak $H$ ($3-5$ mT). The second group (II) are the weak supercurrents ($1-2 \;\mu$A) seen in the oscillations in $(dV/dI)_0$ which survive to fields up to 90 mT (Figs. \ref{figcentral} and \ref{fignoise}).

By tailoring the contact designs in 8 devices, we have found that the group II features are specific to edge states. The oscillation period corresponds to a fluxoid area ${\cal A}_\phi$ bracketed by the Nb contacts (Sec. I in SM). Its perimeter is comprised of two segments (length $w$) lying under Nb contacts, and two segments called ``links'' (of length $\ell_1$ and $\ell_2$) at the uncovered physical edges (inset in Fig. \ref{figcentral}a).

In five devices ($SA,SK,SI,SJ,SB$) the Nb contacts are laid down as parallel strips with spacing $d$ (Table S1, Fig. \ref{figsamples} and Fig. S3 in SM). In the remaining three ($SF,SH,SD$), the Nb contacts are progressively reduced in volume while $\ell_1$ and $\ell_2$ are greatly lengthened. These changes strongly attenuate group II features.

In Fig. \ref{figsamples}, we show two parallel-strip devices, $SA$ and $SI$ (Panels a and b, respectively), and two devices, $SH$ and $SD$, in the second series (c and d, respectively).
The corresponding $(dV/dI)_0$ curves are displayed in Figs. \ref{figsamples}e-\ref{figsamples}h, together with the color maps in Figs. \ref{figsamples}i-\ref{figsamples}j. Group II features are prominently seen in $SA$ and $SI$ where we have $2w\gg \ell_1=\ell_2= d$. As we increase $d$, the widths of the antihysteretic loops decrease. In devices $SH$ (with $\ell_1$ = 5,420 and $\ell_2$ = 5,710 nm) and $SD$ ($\ell_1$ = 8,030 nm and $\ell_2$ = 17,670 nm), the edge oscillations and antihystereses are completely suppressed (aside from residual wings in $SH$). Data from 4 other samples are shown in Fig. S3 in SM.

In devices $SA$ and $SI$, supercurrent in the bulk flows in parallel with supercurrent in the links if $|H|$ is weak. When $|H|$ exceeds $\sim$4 mT, the bulk supercurrent is suppressed but the link supercurrents survive to large $H$ because they are one-dimensional. The robustness enables fluxoid oscillations to be observed up to 90 mT. To realize this, however, we need $\ell_1$ and $\ell_2$ to be shorter than the decay length $\lambda$ ($\sim 800$ nm) that governs the decay of the pair amplitude $\eta$ along a link (inset in Fig. \ref{figcentral}a). If $\ell_1$, $\ell_2$ $>\lambda$ (the case in SH and SD), $\eta$ decays to zero before the loop is closed, and the oscillations vanish (Fig. \ref{figsamples}g,h). The vanishing of the closed-loop supercurrent is also evident in $(dV/dI)_0$ vs. $H$, which stays finite except when $|H|<$ 2 mT. These findings establish that the group II features arise from edge supercurrents at the links.

By comparison, the much larger $I_c$ values of the group I supercurrents (central peak in Fig. \ref{figcentral}a) suggest a bulk origin. The 20-fold decrease in the peak value $I_{\rm c,max}$ as $d$ increases from 156 to 700 nm also implies bulk states proximitized by $\eta$ (Table S1 in SM). The decrease is consistent with the short coherence length of Nb ($\xi_0\sim$ 40 nm). By contrast, at the links, $\lambda\gg\xi_0$.

\vspace{3mm}\noindent
\emph{Proximity Effect Between Competing Symmetries}\\\noindent
In de Gennes' treatment of the proximity effect (valid in the linear-gap regime) the amplitude for a Cooper pair to propagate between points ${\bf r}'$ and ${\bf r}$ is given by the kernel $K({\bf r,r}')$~\cite{deGennes,deGennesDeutscher,Sigrist} (Sec. VII in SM). Whenever a propagator segment lies in Nb, $K({\bf r,r}')$ gains a large enhancement from the strong $s$-channel attraction. Hence the gap function $\hat\Psi({\bf r})$ in MoTe$_2$ can adopt either the symmetry of $F_{\alpha\beta}$ or that of $\eta$ depending on the weighted contributions of propagators going from all points $\bf r'\to\bf r$ (Eq. S23 in SM). An applied field $H$ can tip the balance by selectively weakening one or the other condensation amplitude. Calculations beyond the linear-gap regime show how the gap function changes symmetry in a junction between an $s$- and a $p$-wave superconductor~\cite{Rainer}.

At $H=0$, the weighted average favors $F_{\alpha\beta}$. Vortices inserted by $H$ initially form a vortex solid. With increasing $H$, melting of the vortex solid leads to a dissipative vortex liquid in which the gap modulus $|\hat\Psi|$ remains finite but its phase fluctuates strongly due to vortex motion. In the vortex liquid, which survives to 80 mT, the loss of long-range phase coherence substantially weakens the pair amplitude $F_{\alpha\beta}$ relative to $\eta$. Conversely, solidification of the vortex system strongly tilts the competition towards the intrinsic $F_{\alpha\beta}$. 

\vspace{3mm}\noindent
\emph{Correlating central peak, phase noise and antihysteresis}\\\noindent
In each device, edge-state pairing at the links may be driven either by $F_{\alpha\beta}$ or $\eta$ from the nearest Nb contact. From the phase noise analysis, we infer that compatibility between edge-pairing and bulk-pairing produces oscillations that are nearly noise-free. By contrast, incompatibility between $s$-wave pairing at the edge and $F_{\alpha\beta}$ in the bulk generates substantial phase noise.

\begin{figure*}[t]
\includegraphics[width=18 cm]{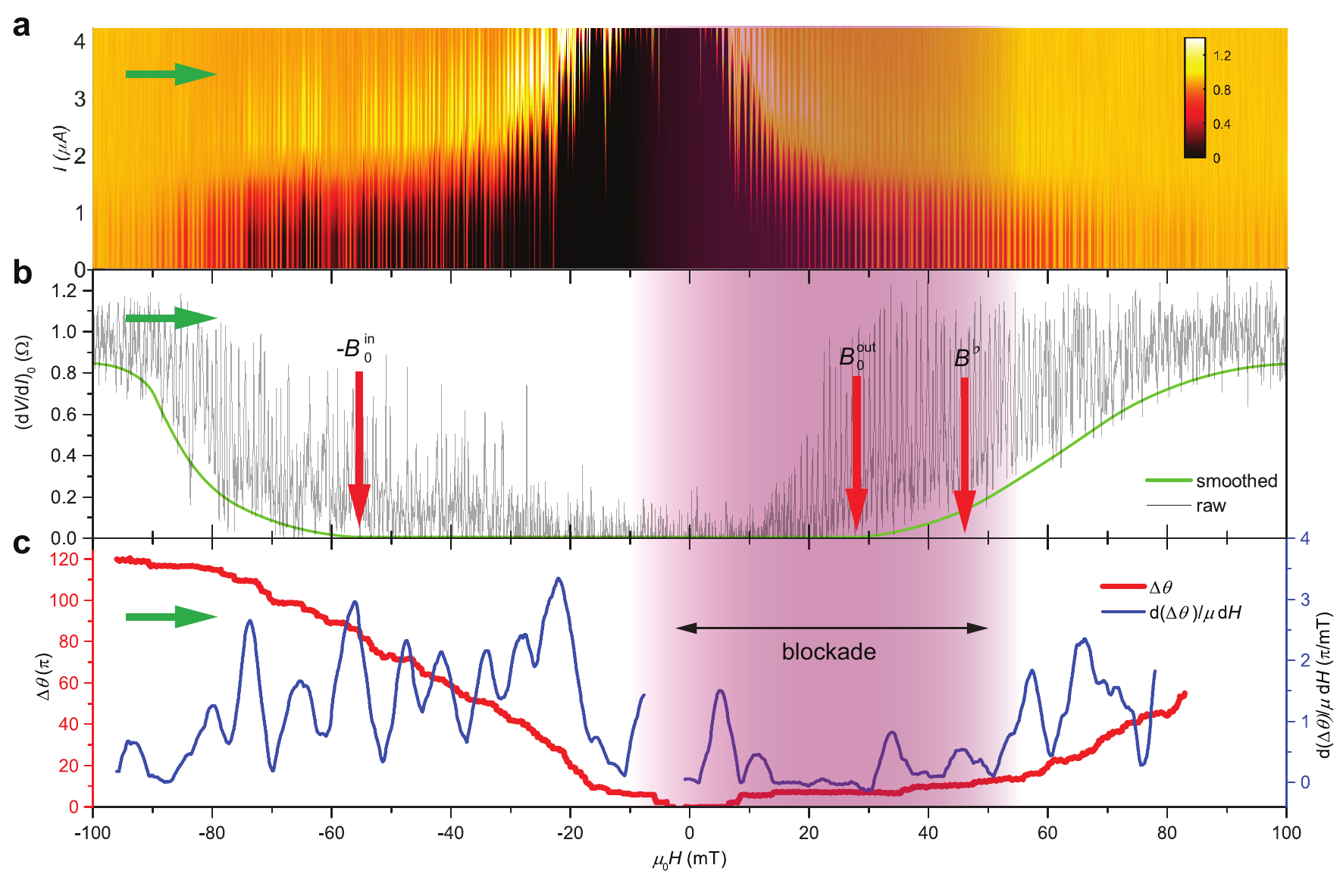}
\caption{\label{figblockade} Correlating the central peak, phase noise and antihysteresis measured in device $SA$ as $H$ is swept from $-100$ to $+100$ mT at 18 mK.
Panel (a) shows the color map of $dV/dI$, highlighting the low-$I$ region. Panel (b) plots the oscillations in the zero-bias $(dV/dI)_0$ (grey curve) together with its floor value (thick green curve). The dissipationless interval for $(dV/dI)_0$ lies between $-B^{\rm in}_0 = -54$ mT and $B^{\rm out}_0 = 28$ mT (red arrows). Panel (c) plots the phase noise measured by $|\Delta\theta|$ (red curve) and $\langle d\Delta\theta/dH\rangle$ (blue). Within the quiet zone extending to $B^\flat$ on the outbound branch (shaded lilac), the phase noise is suppressed. We infer the existence of a blockade mechanism in this interval. In each panel the green arrow indicates the field-scan direction.
}
\end{figure*}

Combining the group I and II features with the phase noise measured by $\Delta\theta$ and $d\langle\Delta\theta/dH\rangle$ in device $SA$, we arrive at a consistent picture of the antihysteretic behavior. The color map of $dV/dI$ and the trace of $(dV/dI)_0$ at 18 mK are displayed in Figs. \ref{figblockade}a and b, respectively. Starting at $-100$ mT, the phase noise is initially small, as seen in $\langle d\Delta\theta/dH\rangle$ (blue curve in Fig. \ref{figblockade}c). This implies that edge pairing is driven by the Nb contacts while all pairing of bulk states is strongly suppressed. When $H$ reaches $-80$ mT, we observe a surge onset in phase noise which we take as evidence for incompatibility with the vortex liquid that appears at $-80$ mT. The phase noise remains large over the entire interval ($-80$, $-12$) mT (red curve in \ref{figblockade}c).

When $H$ crosses $-5$ mT, proximitization of the bulk states driven by $\eta$ becomes strongly favored ($H$ becomes too weak to depair the bulk $s$-wave condensate). The abrupt emergence at $-4$ mT of a large bulk supercurrent (the central peak) signals when $\hat\Psi$ switches to $s$-wave pairing (Figs. \ref{figblockade}a), but this favorable interval is brief. When $H$ reaches $-2$ mT, solidification of the vortex liquid (and the ensuing long-range phase coherence) cause $\hat\Psi({\bf r})$ to switch back to the intrinsic pairing symmetry $F_{\alpha\beta}$. Suppression of all $s$-wave paired regions throughout the crystal collapses the central peak \emph{before} $H$ reaches 0$^-$. Hence the placement of the central peak is antihysteretic as observed.

\vspace{3mm}\noindent
\emph{Blockade}\\\noindent
When the field crosses zero to the outbound branch ($H>0$), we should expect to see the re-appearance of the central peak in the interval (2, 4) mT. However, this is not observed in any of the 8 devices studied. 

Together, the absence of the central peak for $H>0$ and the noise suppression in the interval $(5,45)$ mT suggest a ``blockade'' mechanism. Once $\hat{\Psi}$ switches to the intrinsic symmetry $F_{\alpha\beta}$ in the limit $H\to 0^{-}$, the intrinsic condensate appears to block re-proximitization of the bulk by $\eta$ on the outbound branch throughout the quiet zone which extends to a field that we call $B^\flat\sim$ 45 mT (shaded lilac in Figs. \ref{figblockade}a-c). Edge pairing by Nb is blocked within the quiet zone.

The identification of $(dV/dI)_0$ with supercurrents at the edge clarifies considerably the anti-hysteretic loops in Figs. \ref{figsamples}e and \ref{figsamples}f. As shown by the green curve in Fig. \ref{figblockade}b, the dissipationless interval for the edge current spans the interval $(-B^{\rm in}_0, B^{\rm out}_0) = (-54, 28)$ mT for a left-to-right field scan. The interval $(-B^{\rm in}_0, B^{\rm out}_0)$ is shifted left because, on the inbound branch, the edges are proximitized by the strong Nb pair potential, whereas, on the outbound, edge pairing is driven by the much weaker $F_{\alpha\beta}$ while $\eta$-pairing is excluded by the blockade. Likewise, in a right-to-left scan, $(-B^{\rm out}_0,B^{\rm in}_0)$ is shifted right. This accounts for the anti-hysteretic sign of the loops in $(dV/dI)_0$.

\begin{figure*}[t]
\includegraphics[width=18 cm]{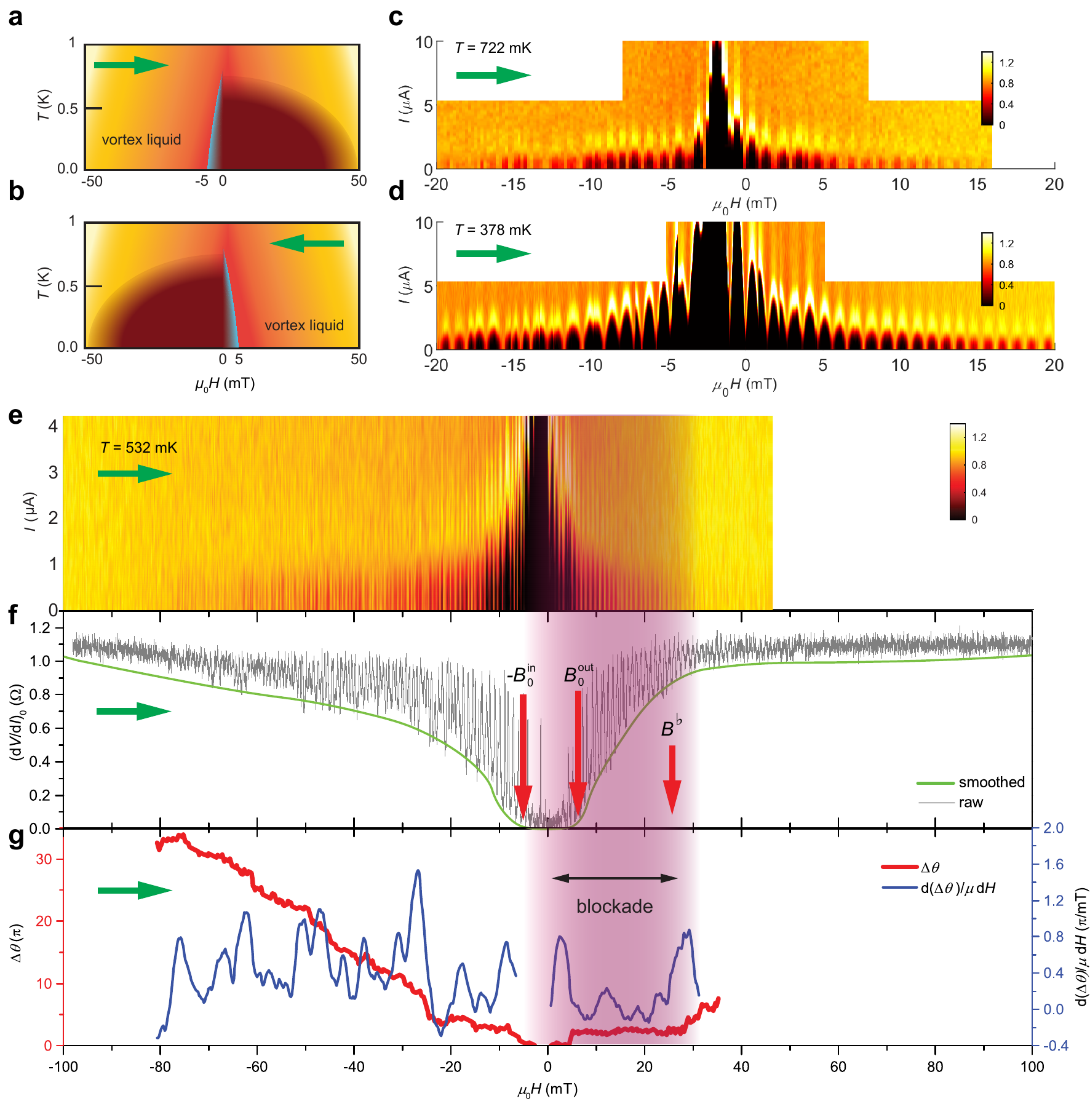}
\caption{\label{fighigh}
Color maps, zero-bias $(dV/dI)_0$ and phase noise at elevated temperatures in device $SA$.
Panels (a) and (b) display the inferred metastable states in the $H$-$T$ plane, for left-to-right and right-to-left field scans, respectively. The regions where the blockade mechanism operates are shaded maroon. Thin blue stripes indicate regions in which the central peak emerges (note the antihysteretic placement). On the outbound branch in (a), the maroon region extends to $B^\flat$ ($\sim 45$ mT) at 18 mK. The dissipative vortex liquid (orange and yellow regions) extend to $\pm 80$ mT. The device $T_c$ is 850 mK. Panels (c), (d) and (e) display color maps of $dV/dI$ measured at 722, 378 and 532 mK, respectively. Panel (f) plots the oscillations in the zero-bias $(dV/dI)_0$ (grey curve) and its floor value (green curve). Panel (g) display the phase noise measured by $\Delta\theta$ (red curve) and  $\langle d\Delta\theta/dH\rangle$ (blue). The lilac regions in Panels (f) and (g) represent the quiet zone in which phase noise is minimal. As $T\to T_c$, the field scales $B^{\rm in}_0$, $B^{\rm out}_0$ and $B^\flat$ decrease to zero. 
In all panels, green arrows indicate the field-scan direction.
}
\end{figure*}

\vspace{3mm}\noindent
\emph{Raising the temperature}\\\noindent
Figure \ref{fighigh}a shows the blockade region (shaded maroon) inferred from left-to-right field scans taken at elevated temperatures 378, 532 and 722 mK (Fig. \ref{fighigh}b is for the opposite field scan). The corresponding color maps of $dV/dI$ are displayed in Figs. \ref{fighigh}c, d and e. For the results at 522 mK, we also show the trace of $(dV/dI)_0$ (Fig. \ref{fighigh}f) and the phase noise measured by $\Delta\theta$ and $\langle d\Delta\theta/dH\rangle$ (Fig. \ref{fighigh}g). The overall patterns are similar to those taken at 18 mK (Fig. \ref{figblockade}) except that raising $T$ decreases the field scales. The asymmetry of the phase noise, including its suppression in the blockade interval, is apparent in Figs. \ref{fighigh}f and \ref{fighigh}g. We find that the widths of the antihysteretic loops in $(dV/dI)_0$, the edge dissipationless interval $(-B^{\rm in}_0, B^{\rm out}_0)$ and $B^\flat$ all decrease systematically as $T$ increases, reaching zero at the device $T_c\simeq 850$ mK. The linear decrease of $B^\flat(T)$ as $T\to T_c$ is shown in Fig. S9. These trends are consistent with the key role played by $F_{\alpha\beta}$ in generating the antihysteresis, phase noise and the blockade.

\vspace{3mm}\noindent
\emph{Hysteretic solid-to-liquid transition}\\\noindent
If we inspect the blockade region in the $H$-$T$ plane (shaded maroon in Figs. \ref{fighigh}a,b), we find that its placement mimics that of a vortex solid phase that is field-shifted by hysteresis. We conjecture that the blockade mechanism and quiet zone are features intrinsic to the vortex solid. If $H$ is swept from left to right (Fig. \ref{fighigh}a), the liquid-to-solid transition is ``delayed'' on the inbound branch until $H$ reaches $-2$ mT, analogous to supercooling of a liquid (regarding $|H|$ as an effective temperature). On the outbound branch, the solid-to-liquid transition is also delayed and shifted to 45 mT. Hence the observed vortex solid phase is displaced to the right as shown in Fig. \ref{fighigh}a. For the opposite scan, the shift is to the left. We note that the supercooling hysteresis has the conventional sign. However, the blockade mechanism and incompatibility between condensates together displace the bulk and edge supercurrent responses in the opposite direction, which leads to the antihysteresis.


\vspace{1cm}
\centerline{* ~~~ * ~~~  *}

\newpage
\vspace{1cm}\noindent
$^{\S}$Corresponding author email: npo@princeton.edu\\

\vspace{5mm}\noindent
{\bf Data availability} \\
The data in the plots will be uploaded to Harvard DataVerse.

\vspace{5mm}\noindent
{\bf Acknowledgements} \\
We have benefitted from discussions with Wudi Wang and Zheyi Zhu. N.P.O. and S.K. were supported by the U.S. Department of Energy (DE-SC0017863). 
The crystal growth effort was supported by a MRSEC grant from the U.S. National Science Foundation (NSF DMR-2011750) which supported S.L, L.M.S., R.J.C and N.P.O. The Gordon and Betty Moore Foundation's EPiQS initiative provided generous support via grants GBMF9466 (N.P.O.) and GBMF9064 (L.M.S.).

\vspace{3mm}
\noindent
{\bf Author contributions}\\
S.K. and N.P.O. conceptualized and designed the experiment. S.K. performed all device fabrication and measurements. The crystals were grown by S.L., L.M.S. and R.J.C. Analysis of the data were carried out by S.K. and N.P.O. who jointly wrote the manuscript.

\vspace{3mm}
\noindent
{\bf Competing financial interests}\\
The authors declare no competing financial interests.

\vspace{3mm}
\noindent
{\bf Additional Information}\\
Supplementary Materials is available in the online version of the paper.

\vspace{3mm}
\noindent
{\bf Correspondence and requests for materials}
should be addressed to N.P.O.


\newpage


{\bf Methods}

\section{Device fabrication}
The MoTe$_2$ devices were fabricated following standard nanofabrication procedures. Substrates with pre-patterned alignment makers were prepared in advance. The alignment markers were made of 5 nm of Ti and 50 nm Au. MoTe$_2$ microflakes were mechanically exfoliated on to the substrates using Nitto Denko SPV 5057A5 Surface Protection Tape. Microflakes with adequate sizes with sharp edges were chosen using an optical microscope.

Niobium electrodes were made following the standard PMMA-copolymer bilayer recipe. After the resists were spun on top of substrates, contacts were patterned using EBPG 5150 from Raith. The devices were developed with MIBK-IPA (1:3) solution for 60 seconds and rinsed with IPA solution. The devices were then transferred to an EvoVac Sputtering System (Angstrom Engineering). An \emph{in-situ} Ar plasma beam bombarded the surfaces of the devices for 120 seconds to remove the top few layers of MoTe$_2$ that were oxidized. A thin layer (3 nm) of Ti layer was sputtered at a rate $0.1 - 0.2$ nm/s to form a sticking layer, followed by 100 nm of Nb deposited at a rate 1 nm/s. Finally, a 5 nm of Au was sputtered on top of Nb layer to protect it from oxidization. All sputtering procedures were performed at a high vacuum of approximately $10^{-8}$ mbar.

\section{Measurement techniques}
\label{sec:methods}

Measurements were carried out in a top-loading dilution refrigerator (Oxford Instruments Kelvinox TLM). Once loaded to the fridge, devices were immersed in the circulating $^3$He-$^4$He mixture in the mixing chamber. The base temperature of the fridge was $T \sim 20$ mK. Three filters were used to reduce noises during measurements. An $LC$ $\pi$-filter, located at the top of the probe, provided 80-dB suppression for frequencies $f > 100$ MHz. The two other filters were located in the mixing chamber. The first one was a sintered metal powder filter that consisted of Cu particles of 30-60 $\mu m$ in diameter. It suppressed any stray electromagnetic radiation for frequencies $f > 1$ GHz. The second filter was a low-pass $RC$ filter with the cutoff frequency of $f = 300$ kHz. A NbTi superconducting magnet was used to apply magnetic fields to devices. A Keithley 2400 and a Keithley 6221 provided the current to superconducting magnet. The smallest field-step size was as small as a few $\mu$T.

Differential resistances of all devices were measured using a SR830 lock-in amplifier. A DC bias from a Keithley 2400 voltage source was superposed with a small AC excitation voltage from the lock-in amplifier through a home-made voltage adder. The resulting voltage was converted to current through a buffer resistor that was orders of magnitude larger in resistance than the device of interest. The voltage signal $V$ across the device first passed through a pre-amplifier to improve the signal-to-noise ratio. The gain was usually in the order of $G \sim 1000$. The amplified signal then reached the lock-in amplifier for detection. Measurements were done in quasi-4-point fashion. Although both voltage and current contacts shared the same electrodes, the electrodes were superconducting at 20 mK.

The differential resistance plots in the main text were obtained through the following procedures: first, the magnetic field was set to the desired value. Then the DC bias was ramped from zero to the desired bias value with a small step size. After reaching the desired bias, the current was ramped back to zero with a much larger step size. Such procedures were repeated for desired field ranges. The collected differential resistance curves were plotted together. The differential resistance versus field at zero bias $(dV/dI)_0$ plots were prepared as follows: starting with the field at zero, we swept $H$ to its maximum negative value then to the maximum positive value before terminating at zero.




\end{document}